\documentclass[a4paper,12pt]{article}
\usepackage[cp1251]{inputenc}
\usepackage{graphicx}
\title{Effect of correlated disorder on the temperature of
 unconventional Cooper pairing ($^3$He in aerogel).}
\author {I. A. Fomin\\
{\it P. L. Kapitza Institute for Physical Problems},\\{\it
Kosygina 2,
 119334 Moscow, Russia}}

\date{ }
\begin{document}
\maketitle

\begin{abstract}
Using liquid $^3$He in aerogel as an example, it is shown that
correlations in positions of impurities affect the temperature
$T_c$ of transition of Fermi liquid in an unconventional
superfluid or superconductive state. The effect is significant if
the correlation length for impurities is greater than the
coherence length in the superfluid or superconductive state
$\xi_0$. For $^3$He in aerogel the suppression of $T_c$ is
expressed in terms of the structure factor of aerogel. With the
account of the fractal structure of aerogel a simple expression is
obtained for the decrease of $T_c$ from its clean value. This
expression is in a satisfactory agreement with the experimental
data.


\end{abstract}

 {\bf 1.} For unconventional Cooper pairing the temperature of
 transition into the superfluid (or superconductive) state is
 suppressed both by magnetic and by non-magnetic impurities
 \cite{Lark}. The amount of suppression in most cases is well
 described by the theory of superconductive alloys of Abrikosov
 and Gor`kov (AG in what follows) \cite{AG}. Superfluid $^3$He is
 the best understood example of unconventional Cooper pairing. The
 Cooper pairs here have orbital momentum $l=1$, and spin $s=1$.
 Floating impurities can not be introduced in the liquid $^3$He --
 they stick to the walls. To get around the difficulty a high-porosity
 silica aerogel has been used as an impurity \cite{parp1,halp}.
 The aerogel is a self-supporting network of silica strands with a
 typical thickness 3-4 nm. Rigidity of the network assumes
 existence of correlations in positions of its elements. The
 homogeneous scattering model (HSM) \cite{thuneb}, which is a
 generalization of the AG theory for the $p$-wave pairing, does not
 take into account these correlations. This drawback is
 essential, since HSM does not provide a quantitative description
 of properties of superfluid phases of $^3$He in a presence of aerogel.
 In particular, HSM does not describe correctly the dependence on pressure
 of the magnitude of suppression of superfluid transition
 temperature of  $^3$He in aerogel $T_a$ with respect to its bulk transition
 temperature $T_b$.

 In the AG-theory, as well as in the HSM, the relative value of the
 suppression $\tau_{ba}={\frac{T_b-T_a}{T_b}}$  is determined by
 the only parameter $x=\xi_0/l_{tr}$, where $\xi_0=\hbar v_F/(2\pi
 T_a)$ is the superfluid coherence length and $l_{tr}$ is a
 transport mean free path. For $\tau_{ba}\ll 1$ the theory predicts
 $\tau_{ba}=\frac{\pi^2}{4}
 \frac{\xi_0}{l_{tr}}$. Both $\xi_0$ and $l_{tr}$ can be found
 from independent experiments \cite{sauls1}. The measured
 transition temperature $T_a$ is greater than the value, calculated
 within the HSM with the use of the known values of $\xi_0$ and $l_{tr}$.
The other suggested models \cite{thuneb,han} invoke an effect of
restricted geometry as a mechanism of suppression of $T_a$. One of
the models considers $^3$He in a gap between two diffusely
scattering planes, and the other -- $^3$He in a spherical void.
Although a better agreement with experiment can be achieved in
this way, relation of the models to the real aerogel remains
unclear.

In a present paper effect of correlation in position of elements,
forming aerogel, on the superfluid transition temperature of
$^3$He is considered directly.  The effect is the stronger the
greater is the ratio of  correlation length in aerogel $R$ to
$\xi_0$. When $R^2\sim l_{tr}\xi_0$  the change of $\tau_{ba}$
stemming from the correlations can be of the order of the original
$\tau_{ba}$. This region is of particular interest for the
aerogels, used in experiments. To simplify the argument we
consider here only a region $\tau_{ba}\ll 1$.

{\bf 2.} At the $p$-wave Cooper pairing the order parameter is
3$\times$3 complex matrix $A_{\mu j}$. The first index assumes
tree values enumerating three projections of spin of the Cooper
pair. The second index enumerates three projections of its orbital
momentum. To find $T_a$ assuming $\tau_{ba}\ll 1$ one can use
Ginzburg and Landau equation:
 $$
-\tau A_{\mu
 j}+A_{\mu l}\eta_{lj}(\mathbf{r})-\frac{3}{5}\xi^2_s\left(\frac{\partial^2 A_{\mu j}}{\partial x_l^2}+
2\frac{\partial^2 A_{\mu l}}{\partial x_l \partial x_j}\right)=0 .
\eqno(1)
$$
Here $\tau=(T_b-T)/T_b\ll 1$. Effect of impurities (aerogel) is
introduced via a real symmetric tensor $\eta_{lj}(\mathbf{r})$, it
describes a local depression of the transition temperature and its
possible splitting for different orbital components of $A_{\mu
j}$. Interaction with the impurities can be written in Eq.(1) in a
local form because tensor $\eta_{lj}(\mathbf{r})$ varies on a
distance $\sim\xi_0$, while  $A_{\mu j}$  in a vicinity of $T_b$
varies on a distance $\sim\xi(T)=\xi_0/\sqrt{\tau}\gg\xi_0$. A
coefficient in front of the derivatives is written as in the Ref.
\cite{ambega}, i.e. $\xi_s^2=\frac{7\zeta(3)}{12}\xi_0^2\simeq
0,7\xi_0^2$.  At $\tau>0$ Eq.(1) can have physically meaningful
solutions. These solutions are generally speaking nonuniform and
can be localized. By the definition $T_a$ is a temperature of the
onset of the long-range order in liquid  $^3$He. In the present
formulation it corresponds to the minimum  $\tau=\tau_a$, at which
Eq.(1) can have delocalized solution. The equations (1) for
different spin components $\mu$ are not coupled and the spin index
is  not relevant. A problem of finding  $\tau_{ab}$ turns out to
be analogous to a problem of finding mobility edge of a spin-1
particle in a random potential  $\eta_{lj}(\mathbf{r})$, and
$\tau$ is analog of energy. To avoid cumbersome calculations let
us introduce one more simplification. Instead of the sum of two
gradient terms in Eq.(1) we use a model isotropic expression with
only one term and an ``average" coefficient $\overline{\xi_s^2}$:
 $$
-\tau A_{\mu
 j}+A_{\mu l}\eta_{lj}(\mathbf{r})-
 \overline{\xi_s^2}\left(\frac{\partial^2 A_{\mu j}}{\partial x_l^2}
\right)=0. \eqno(2)
$$
The average $\overline{\xi_s^2}$ is defined in the following way.
For the longitudinal component of $A_{\mu l}$, which meets the
condition $\frac{\partial A_{\mu l}}{\partial x_l}=0$, the
coefficient in front of the derivative in Eq.(1) is
$(3/5)\xi_s^2$. For two transverse components, defined as  $A_{\mu
l}=\frac{\partial\psi_{\mu}}{\partial x_l}$, where $\psi_{\mu}$
are scalars, a proper coefficient is $(9/5)\xi_s^2$. The averaging
over three possibilities renders
$\overline{\xi_s^2}=(7/5)\xi_s^2$. Since $\xi_s^2\simeq
0,7\xi_0^2$ with a reasonable accuracy  $\overline{\xi_s^2}\simeq
\xi_0^2$. In what follows $\xi_0^2$ will be used as a coefficient
in front of the gradient term in Eq.(2).

Explicit form of tensor $\eta_{lj}(\mathbf{r})$ depends on the
particular structure of aerogel. We assume here, that aerogel
consists of spheres of uniform radii $\rho$, distributed with an
average density $n$. The values of $\rho$ and $n$ are chosen to
provide the required values of porosity and of the mean free path
for single-particle excitations. It will became clear later, that
a concrete form of the elements is not essential, but for explicit
calculations spherical form is preferable.  For the porosity
98,3\% and $l_{tr}\approx$130 nm $\rho\approx$2 nm, i.e.
 $\rho\ll\xi_0$.
 For such model
$$
\eta_{jl}({\bf r})=\sum_s\eta_{jl}^{(1)}({\bf r-r_s}), \eqno(3)
$$
where ${\bf r_s}$ is a coordinate of the sphere number $s$, and
$\eta_{jl}^{(1)}({\bf r})$ a potential induced by a single sphere,
it depends on the cross-section of scattering of the
single-particle excitations on the sphere \cite{Rainer}. Assuming
for the sake of definiteness diffuse scattering, we arrive at
 $r\gg\rho$ at the expression
\cite{fom_j}:
$$
\eta_{jl}^{(1)}({\bf
r})=-\frac{\rho^2}{r^2}\hat\nu_j\hat\nu_l\ln\left[
\tanh\left(\frac{r}{2\xi_0}\right)\right]. \eqno(4)
$$

According to Eq.(4) $\eta_{jl}^{(1)}({\bf r})$ decays on a
distance $\sim\xi_0$. At $r\leq\xi_0$ $\eta_{jl}^{(1)}({\bf
r})\sim(\rho/\xi_0)^2$. Let us treat $\eta_{jl}({\bf r})$ as a
perturbation.  Green function for Eq.(2) after averaging over
realizations of $\eta_{jl}({\bf r})$ has a form: $\langle
G_{mn}(\tau;{\bf k},{\bf k'})\rangle=(2\pi)^3\delta({\bf k}-{\bf
k'})\delta_{mn}G(\tau;{\bf k})$, where
$$
G(\tau;{\bf k})=\frac{1}{\tau-\xi_0^2k^2-\Sigma(\tau,{\bf k})}.
\eqno(5)
$$
The self-energy $\Sigma(\tau,{\bf k})$ up to the factor
$(2\pi)^3\delta({\bf k}-{\bf k'})\delta_{mn}$ is the averaged sum
of the series shown diagrammatically in Fig.1.
\begin{figure}
\begin{center}
\includegraphics[%
  width=0.75\linewidth,
  keepaspectratio]{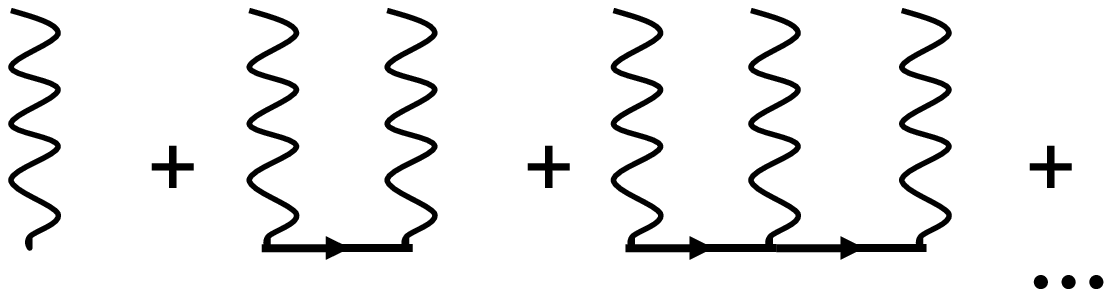} 
\end{center}
\caption {}
\end{figure}
The arrows in the figure correspond to the unperturbed Green
functions
$$
G^{(0)}_{mn}(\tau;{\bf k})=\frac{\delta_{mn}}{\tau-\xi_0^2k^2}.
\eqno(6)
$$
As usual, the integration over momenta of internal lines is
assumed. Wavy lines correspond to the Fourier transform of the
potential $\eta_{jl}({\bf r})$:
$$
\eta_{jl}({\bf k-k'})=\eta_{jl}^{(1)}({\bf k-k'})\sum_se^{i({\bf
k'}-{\bf k}){\bf r}_s},                                \eqno(7)
$$
where {\bf k} and  {\bf k}$'$ - momenta, corresponding to the
arrows coming in and out of the vertex. The averaging is performed
over coordinates  ${\bf r}_s$ of particles forming aerogel. The
``mobility edge" is found as a pole of the Green function Eq.(5)
at $k=0$:
$$
\tau=\Sigma(\tau,0).              \eqno(8)
$$
Fourier transform of $\eta_{jl}^{(1)}({\bf k-k'})$ can be found
directly from Eq.(4). Here we need only  $\eta_{jl}^{(1)}({\bf
k}\rightarrow 0)=\delta_{jl}\eta^{(1)}(0)$, where:
$$
\eta^{(1)}(0)=\frac{\pi^3}{3}\rho^2\xi_0.          \eqno(9)
$$
Consider consecutive terms of the series Fig.1. The first order
term is: $$ \langle\eta_{jl}({\bf k}-{\bf
k'})\rangle=\eta_{jl}^{(1)}({\bf k}-{\bf
k'})\langle\sum_se^{i({\bf k'}-{\bf k}){\bf r}_s}\rangle.
  \eqno(10)
$$
Assuming that aerogel is on the average uniform we have:
$$
\langle\sum_se^{i({\bf k'}-{\bf k}){\bf
r}_s}\rangle=(2\pi)^3\delta({\bf k}-{\bf k'})n.      \eqno(11)
$$
As a result, in the first order on the perturbation
$$
\tau_{ba}^{(1)}=n\eta^{(1)}(0)=\frac{\pi^2}{4}\frac{\xi_0}{l_{tr}}.
\eqno(12)
$$
For the comparison with HSM the answer is expressed in terms of
the  transport mean free path of the excitations $l_{tr}$, which
is defined as:
$$
\frac{1}{l_{tr}}=\frac{4}{3}\pi\rho^2n.  \eqno(13)
$$
This definition corresponds to a diffuse scattering of
quasi-particles by the randomly distributed uniform spheres with
radii $\rho$. The first order correction (12) coincides with the
result of HSM for small $\tau_{ba}$. The second-order correction
is:
$$
\Sigma^{(2)}(\tau,{\bf k})\delta_{jl}=\int\eta^{(1)}_{jm}({\bf
k}-{\bf k_1})\eta^{(1)}_{nl}({\bf k_1}-{\bf
k})n\langle\sum_{t}e^{i({\bf k_1}-{\bf k}){\bf r}_{st}}\rangle
G_{mn}(\tau,{\bf k_1})\frac{d^3k_1}{(2\pi)^3}. \eqno(14)
$$
Instead of the unperturbed Green function Eq.(6) the average Green
function Eq.(5) is substituted in the r.h.s. of Eq.(14). This
substitution is known as the self-consistent Born approximation
\cite{varma}. The summation in Eq. (14) is over the relative
coordinate ${\bf r}_{st}={\bf r}_s-{\bf r}_t$. The averaged sum is
the structure factor:
$$
\langle\sum_{t}e^{i({\bf k_1}-{\bf k}){\bf r}_{st}}\rangle\equiv
S({\bf k_1}-{\bf k}),  \eqno(15)
$$
it characterizes correlations in positions of the particles,
forming aerogel. A structure factor is directly measured by the
small-angle x-ray scattering. For the isotropic aerogel $S({\bf
k})$ does not depend on a direction of ${\bf k}$. Substituting
 ${\bf k}=0$ in Eq. (14) we arrive at:
$$
\tau_{ba}^{(2)}\delta_{jl}=n\int\eta^{(1)}_{jm}(-{\bf k_1})
\eta^{(1)}_{ml}({\bf k_1})
S(k_1)G(\tau_{ba},k_1)\frac{d^3k_1}{(2\pi)^3}. \eqno(16)
$$
Following Eq. (8) and neglecting a possible small change of the
spectrum we arrive at $G(\tau_{ba},k_1)=-1/(\xi_0k_1)^2$. At
evaluation of the integral in Eq.(16) an account must be taken of
the fact that in an interval of scales $\varrho<r<R$, where $R$ is
 an upper limit of the interval, aerogel has fractal structure.
It means, that in the corresponding interval of wave vectors
$k_{min}<k<k_{max}$ its structure factor $S$ has a power
dependence on $k$: $S\sim k^{-D}$. The exponent $D$ is referred as
a fractal dimension. In particular for the samples of 98\% aerogel
used in Ref.\cite{parp2} $k_{min}\simeq 5\div 7\cdot
10^{-3}\AA^{-1}$ , $k_{max}\simeq 1\div 2\cdot 10^{-1}\AA^{-1}$,
$D\approx 1,8$. This rate of decay of $S(k)$ secures convergence
of the integral in the infinity.  A principal contribution to the
integral is provided by the region of small  $k$. That allows to
substitute in the integrand  $\eta^{(1)}_{jm}({\bf k}-{\bf
k_1})\simeq\eta^{(1)}_{jm}(0)=\delta_{jl}\eta^{(1)}(0)$, then
$$
\tau_{ba}^{(2)}=n[\eta^{(1)}(0)]^2\int
S(k_1)G(0,k_1)\frac{d^3k_1}{(2\pi)^3}. \eqno(17)
$$
Therefore a particular shape of the elements, forming aerogel is
not essential. Within the fractal interval the integral in Eq.(17)
is $\sim\int dk/k^D$ and at $k_{min}\to 0$ it diverges at small
$k$.
 Actually the integral converges since at $k<k_{min}$
dependence $S(k)$ saturates. That reflects an absence of
correlations on a distances exceeding $R\sim 1/k_{min}$. A smooth
cut-off can be introduced by assuming a simple model law of decay
of correlations, depending on $R$. Then, by definition, $R$ is a
correlation radius. The structure factor $S({\bf k})$ can be
expressed in terms of the pair correlation function $C({\bf r})$:
$$
S({\bf k})=\langle\sum_{t}e^{i{\bf k}{\bf r}_{st}}\rangle=n\int
C({\bf r})e^{i{\bf k}{\bf r}}d^3r.   \eqno(18)
$$
At  $r\ll R$ for a fractal with the dimension $D$ correlation
function behaves as: $C(r)\approx A(R/r)^{3-D}$, where $A$ - is a
coefficient.  At $r\gg R$ correlations vanish and $C(r)\rightarrow
1$, i.e. the integral in Eq.(18) diverges. It converges if $C(r)$
is substituted by $v(r)=C(r)-1$. This substitution corresponds to
separation of the effect of correlations from that of uncorrelated
distribution of elements. The unity gives a contribution to
$S({\bf k})$, which is proportional to $\delta({\bf k})$ and does
not affect the following calculations. The integral in Eq. (17),
denoted as $Q$ in what follows, with the aid of straightforward
transformations can be expressed in terms of $v(r)$:
$$
Q=\int S(k_1)G(k_1)\frac{d^3k_1}{(2\pi)^3}=-\frac{n}{\xi_0^2}\int
v(r)rdr. \eqno(19)
$$
At $r\ll R$ function $v(r)\sim A(R/r)^{3-D}$. For $D>1$ the
integral in Eq.(19) converges at small $r$, therefore the fractal
asymptotic can be used up to  $r=0$. At $r\geq R$  vanishing of
correlations has to be taken into account. Following Ref.
\cite{kleman} we assume here the exponential decay of
correlations, i.e. substitute for $v(r)$:
$$
v(r)= [A\left(\frac{R}{r}\right)^{3-D}-1]exp(-r/R).   \eqno(20)
$$
Coefficient $A$ is found from the normalization condition:
 4$\pi n\int v(r)r^2dr=-1$.
For aerogels in question  $nR^3\gg 1$. Then
  $A=2/\Gamma(D)$, where $\Gamma(D)$ is Euler Gamma-function.
With this $A$ we arrive at: $\int v(r)rdr=R^2(3-D)/(D-1)$. As a
result
$$
\tau_{ba}^{(2)}=-(n\eta^{(1)}(0))^2\frac{R^2}{\xi_0^2}\frac{3-D}{D-1}=
-\frac{\pi^2}{4}\frac{\xi_0}{l_{tr}}
\left(\frac{\pi^2}{4}\frac{R^2}{\xi_0l_{tr}}\frac{3-D}{D-1}\right).
\eqno(21)
$$
 Suppression of the transition
temperature with the account of the second order term
$$
\tau_{ba}=\frac{\pi^2}{4}\frac{\xi_0}{l_{tr}}-\frac{\pi^2}{4}\frac{\xi_0}{l_{tr}}
\left(\frac{\pi^2}{4}\frac{R^2}{\xi_0l_{tr}}\frac{3-D}{D-1}\right)
\eqno(22)
$$
turns out to be smaller then that, predicted by HSM. The factor

$\frac{\pi^2}{4}\frac{R^2}{\xi_0l_{tr}}\frac{3-D}{D-1}$ grows when
 $\xi_0$ decreases. Compensation of the small parameter $\xi_0/l_{tr}$
by a big ratio $(R/\xi_0)^2$ can result in a product of the order
of unity. In this case higher order terms in the expansion of
$\tau_{ba}$ over $\eta_{jl}$ have to be taken into account as
well. The third order term is:
$$
\tau_{ba}^{(3)}=
n(\eta^{(1)}(0))^3\int\frac{d^3k_1}{(2\pi)^3}\frac{d^3k_2}{(2\pi)^3}
G(k_1)G(k_2)\langle\sum_{t,u} e^{i{\bf k_1}{\bf r}_{st}}e^{i{\bf
k_2}{\bf r}_{us}}\rangle.     \eqno(23)
$$
The averaged sum here depends on three-particle correlations. The
higher order terms respectively depend on the higher order
correlation functions. The sum of the series can be found if the
assumption is made, that the higher order correlation function can
be decoupled in  products of the two-particle correlation
functions. In particular:
$$
\langle\sum_{t,u} e^{i{\bf k_1}{\bf r}_{st}}e^{i{\bf k_2}{\bf
r}_{us}}\rangle=S({\bf k_1})S({\bf k_2}).    \eqno(24)
$$
In this case  $\tau_{ba}^{(3)}$ is a product of  $\tau_{ba}^{(2)}$
by $\eta^{(1)}(0)Q$ etc.. The consecutive terms form geometric
series with the sum:
$$
\tau_{ba}=\frac{\tau^{(1)}}{1-\tau^{(1)}Q}. \eqno(25)
$$
With the given above expressions for  $\tau^{(1)}$ and $Q$ we
arrive at
$$
\tau_{ba}=\frac{\frac{\pi^2}{4}\frac{\xi_0}{l_{tr}}}{1+
\frac{\pi^2}{4}\frac{R^2}{\xi_0l_{tr}}\frac{3-D}{D-1}}. \eqno(26)
$$
The obtained expression for $\tau_{ba}$ in both limits of large
and small $R$ gives physically natural results. At $R\ll\xi_0$ the
result of HSM is reproduced. If, on the other hand,
 $R$  is so large that $\frac{\xi_0}{l_{tr}}\left(\frac{R}{\xi_0}\right)^2\gg 1$,
 then the mean free path  $l_{tr}$ drops out of the expression
 for the transition temperature: $\tau_{ba}\sim\left(\frac{\xi_0}{R}\right)^2$.
 Such behavior is in line with the qualitative
argument, based on the presence in aerogel of the low-density
regions, or `voids', with a characteristic size $\xi_a$
\cite{thuneb,sauls2}. According to the argument for
$\xi_0\leq\xi_a$ superfluidity sets on starting from interior of
the `voids', then $\tau_{ba}\sim(\xi_0/\xi_a)^2$, while at
$\xi_0\gg\xi_a$ homogeneous limit is recovered
$\tau_{ba}\sim\xi_0/l_{tr}$. For interpolation between the two
limits Sauls and Sharma \cite{sauls2} suggested to substitute in
the formula of HSM for $T_a$ instead of the pairbreaking parameter
$x=\xi_0/2l_{tr}$ an effective pairbreaking parameter
$\tilde{x}=x/(1+\zeta_a^2/x)$ with $\zeta_a=\xi_a/l_{tr}$. The
obtained expression turns out to be in a good agreement with the
data for 98\% aerogel. Eq.(26)  can also be rewritten as the HSM
formula (12) with a substitution of the parameter $\tilde{x}$
instead of $x$. To do this one has to set
$\xi_a=R\pi\sqrt{(3-D)/8(D-1)}$ in the definition of $\tilde{x}$.
Therefore Eq.(26) with a proper choice of $R$ has also agree with
the data for $\tau_{ba}$, when $\tau_{ba}$ is small. The
constraint is not crucial. The method of correlation function
\cite{usadel} makes possible generalization of the obtained result
for finite $\tau_{ba}$.

Good agreement with the data for $T_a$ in a 98\% aerogel
\cite{sauls2} is achieved at $\xi_a\approx$500$\AA$ and
$l_{tr}\approx$1400$\AA$. With these values of parameters at
pressure above  $\simeq$20 bar $\zeta_a^2/x\approx$1,6, i.e.
effect of correlation is essential.

In conclusion, one can see that the account of the correlations in
positions of the elements forming aerogel explains suppression of
the temperature of transition of $^3$He in the superfluid state by
aerogel. The observed difference in properties of superfluid
$^3$He in aerogel from the predictions, based on the HSM
\cite{parp3}, indicates that correlations are essential for a
proper interpretation of these properties as well.

The above argument can be applied to unconventional
superconductors as well. In particular, for some of high-T$_c$
superconductors the AG-theory does not properly describe the
suppression of $T_c$ by impurities. The numerical analysis of the
Ref.\cite{franz} demonstrates that the discrepancy originates from
the fact that the correlation length $\xi_0$ in these materials is
significantly smaller than in conventional superconductors and can
be comparable with the average distance between the impurities.

I acknowledge the useful discussion with  A.F. Andreev,  V.I.
Marchenko, A.Ya. Parshin, L.P. Pitaevskii and I.M. Suslov.
 This work is partly supported by RFBR (grant 07-02-00214) and by Ministry of
Science and Education of the Russian Federation.


\begin{thebibliography}{99}

\bibitem{Lark} A. I. Larkin, \emph{ZhETF}, {\bf 58}, 1466 (1970) [Sov. Phys. JETP,
{\bf 31}, 784 (1970)]

\bibitem{AG} A.A. Abrikosov and L.P. Gorkov, \emph{ZhETF}  {\bf 39}, 1781 (1961),
  [\emph{Sov. Phys. JETP} {\bf 12}, 1243 (1961)].

\bibitem{parp1} J. V. Porto and J. M. Parpia, {\it Phys. Rev. Lett.},
               {\bf 74}, 4667 (1995)

\bibitem{halp} D. T. Sprague,T. M. Haard,J. B. Kycia, V. R. Rand, Y. Lee,
P. Hamot and W. P. Halperin, {\it Phys. Rev. Lett.},
               {\bf 75}, 661 (1995)

\bibitem{thuneb} E.V. Thuneberg, S.-K. Yip, M. Fogelstrom, and J.A.
Sauls, {\it Phys. Rev. Lett.}, {\bf 80}, 2861 (1998)

\bibitem{sauls1} J.A. Sauls, Yu. M. Bunkov, E. Kollin, H. Godfrin.
and P. Sharma, {\it Phys. Rev.}, {\bf B72}, 024507 (2005)

\bibitem{han} R. Hanninen and E.V. Thuneberg, {\it Phys. Rev.}, {\bf
B67},214507 (2003)

\bibitem{ambega} V. Ambegaokar, P. G. deGennes, and D. Rainer, {\it Phys. Rev.},
               {\bf A 9}, 2676 (1974)

\bibitem{Rainer} D. Rainer and M. Vuorio, J. Phys. C: Solid State Phys.,
                 {\bf 10} (1977) 3093 .

\bibitem{fom_j} I.A. Fomin, Journ. Phys. and Chem. of Solids, {\bf
66}, 1321 (2005)

\bibitem{varma} S. Schmitt-Rink, C.M. Varma and A.E. Ruckenstein,
{\it Phys. Rev. Lett.}, {\bf 60}, 2793 (1988).

\bibitem{parp2} J. V. Porto and J. M. Parpia, {\it Phys. Rev.},
               {\bf B59}, 14583 (1999).

\bibitem{kleman} T. Freltoft, J.K. Kjems, S.K. Sinha, {\it Phys. Rev.},
{\bf B33}, 269 (1986)

\bibitem{sauls2} J.A. Sauls, and P. Sharma, {\it Phys. Rev.}, {\bf B68},
224502 (2003)

\bibitem{usadel} G. Luders and K.-D. Usadel, The Method of the
Correlation Function in Superconductivity Theory., in Springer
Tracts in Modern Physics, vol. 56, Springer 1971.


\bibitem{parp3} J. M. Parpia, A. D. Fefferman, J.V. Porto, V.V. Dmitriev,
L.V. Levitin, and D.E. Zmeev, {\it J. Low Temp.Phys.}\textbf{150},
464, (2008).

\bibitem{franz} M. Franz, C. Kallin, A.J. Berlinsky, and M.I.
 Salkola, {\it Phys. Rev.}, {\bf B56}, 7882 (1997).

















\end{thebibliography}
\end{document}